\documentclass{emulateapj}
\usepackage{graphicx}
\usepackage{multirow}
\usepackage{textcomp}
\usepackage{epsfig}
\usepackage{amsmath}
\usepackage{amssymb}
\usepackage{amsthm}
\usepackage{xy}

\def\uu {4U\,0142$+$614}
\def\ee {1E\,1048.1$-$5937}
\def\kes {1E\,1841$-$045}
\def\aa {1E\,1547$-$5408}

\def\rxs {1RXS\,J1708$-$4009}
\def\xte{XTE\,J1810$-$197\,}

\def\wes{CXOU\,J1647$-$4552}

\def\ea{1E\,2259$+$586}
\def\sgra{SGR\,1806$-$20}
\def\sgrb{SGR\,1900$+$14}

\def\sgrd{SGR\,1627$-$41\,}
\def\sgre{SGR\,0501$+$4516}
\def\sgrf{SGR\,0418$+$5729}
\def\sgrg{SGR\,1833$-$0832}

\newcommand{\swift}{{\em Swift}}

\newcommand{\bc}{\begin{center}}
\newcommand{\ec}{\end{center}}
\def\ltsima{$\; \buildrel < \over \sim \;$}
\def\lsim{\lower.5ex\hbox{\ltsima}}
\def\loe{\lower.5ex\hbox{\ltsima}}
\def\gtsima{$\; \buildrel > \over \sim \;$}
\def\gsim{\lower.5ex\hbox{\gtsima}}
\def\goe{\lower.5ex\hbox{\gtsima}}

\def\ltsima{$\; \buildrel < \over \sim \;$}
\def\lsim{\lower.5ex\hbox{\ltsima}}
\def\loe{\lower.5ex\hbox{\ltsima}}
\def\gtsima{$\; \buildrel > \over \sim \;$}
\def\gsim{\lower.5ex\hbox{\gtsima}}
\def\goe{\lower.5ex\hbox{\gtsima}}

\def\ergs {erg\,s$^{-1}$}
\def\ergscm2 {erg\,s$^{-1}$cm$^{-2}$}

\def\cm2 {cm$^{-2}$}

\def\ergs {${\rm erg\, s}^{-1}$}

\shorttitle{Modeling magnetar outbursts}
\shortauthors{Pons \& Rea}

\begin{document}

\title{Modeling magnetar outbursts:  Flux enhancements and the connection with short bursts and glitches.}

\author{J. A. Pons\altaffilmark{1} \& N. Rea\altaffilmark{2}}
\altaffiltext{1}{ Department de F\'{\i}sica Aplicada, Universitat d'Alacant, Ap. Correus 99, 03080 Alacant, Spain} 
\altaffiltext{2}{Institut de Ci\`encies de l'Espai (CSIC-IEEC), Campus UAB, Facultat de Ci\`encies, 
Torre C5-parell, E-08193 Barcelona, Spain}

\begin{abstract}

The availability of a large amount of observational data recently collected from magnetar outbursts is now calling for a complete theoretical study of outburst characteristics. 
In this letter (the first of a series dedicated to model magnetar outbursts), we tackle the long-standing 
open issue of whether or not short bursts and glitches are always connected to long-term radiative outbursts. 
We show that the recent detection of short bursts and glitches seemingly unconnected to outbursts 
is only misleading our understanding of these events. 
We show that, in the framework of the starquake model, neutrino emission processes in the magnetar crust limit the temperature, and therefore the luminosity. This natural limit to the maximum luminosity makes outbursts 
associated with bright persistent magnetars barely detectable. These events are simply seen as a small luminosity 
increase over the already bright quiescent state, followed by a fast return
to quiescence. In particular, this is the case for  \rxs, \kes, \sgra, and other bright persistent magnetars. 
On the other hand, a similar event (with the same energetics) in a fainter source will drive a more extreme luminosity variation and longer cooling time, as for sources such as \xte, \aa\, and \sgrd. We conclude that the non-detection of large radiative outbursts in connection with glitches and bursts from bright persistent magnetars is not surprising per se, nor it needs of any revision on the glitches and burst mechanisms as explained by current theoretical models.

\end{abstract}


\section{Introduction}\label{intro}

After the discovery of the first large transient event from  \xte\, \citep{ibra04}, the study of transient long-term activity of Soft Gamma Repeaters (SGRs) and Anomalous X-ray Pulsars (AXPs) has provided a new tool to study the physics of strongly magnetized neutron stars (see \cite{mereghetti08,reaesp11} for recent reviews). At present,  thanks to  wide field monitors such as the {\em Swift} Burst Alert Telescope (BAT) and the {\em Fermi} Gamma-ray Burst Monitor (GBM), we are currently detecting one or two new outbursts per year, both from known or newly discovered sources. The interpretation of the large amount of observational data accumulated leads to (apparently contradictory) conclusions concerning the connection of large, long-term flux variations (outbursts) with the occurrence of short bursts and/or glitches.

Already from one of the first outbursts discovered,  the connection between the occurrence of glitches, short bursts and the increase in the persistent flux of \ea\ was clear \citep{kaspi03,woods04,zhu08}. Not only all three phenomena were 
observed at the same time, but this was also in agreement with a theoretical explanation for how bursts and outbursts
are generated. They are thought to be caused by large scale rearrangements of the crustal and/or magnetospheric field, resulting in  the fracture of the neutron-star crust when magnetic stresses locally exceed the tensile
strength of the crust. The {\it starquake} is accompanied by the release of elastic and magnetic energy and may result in renewed magnetospheric activity and additional hot spots on the neutron-star surface. This is the plausible cause of spectral changes during outbursts, pulse profile variability, and differences in cooling patterns. The energetics and event frequency 
depend not only on the strength of the dipolar field, but also on the intensity
and geometry of the internal field and the age of the source \citep{pons11}.

However, although many sources were discovered through their outbursts, and later linked to the occurrence of short bursts and glitches, in many other cases bursts and glitches occurred without the detection of a simultaneous outburst \citep{dib08}, or with only very subtle flux variations \citep{rea05}. 
This apparently random connection between glitches/bursts and long outbursts prompted further questions related to the theory behind the crustal fractures, and originated ideas related to the possible magnetospheric origin of bursts not connected with large radiative enhancements \citep{lyu06}.
For all these reasons, disentangling the connection between transient outbursts, glitches and short and large bursts, has been one of the major issues in the magnetar field in the past few years. In this letter
we discuss the circumstances under which the simultaneous detection of bursts, glitches and a
long outburst is expected.  At the same time we give an explanation for the apparent lack of connection in some cases. 

%%%%%%%%%%%%%%%%%%%%%%%%%%%%%%%%%
\section{Observational ground}
\label{obs}
In this section we summarize our current knowledge of magnetar flux variations and their connection with bursting and glitching behaviour.  

\subsection{Glitches and/or short bursts with very subtle or no flux variability} 

\kes, embedded in the bright SNR Kes 73, is one of the most prolific glitcher among magnetars \citep{dib08}. It recently showed a few bursts \citep{gav11b}. However, its flux was never observed to vary significantly despite long term monitoring programs with most of the current X-ray satellites \citep{zhu10,lin11}.

Other sources where only very subtle flux changes were observed are \rxs\, and \uu\, (by a factor $<2$). The weak flux variability from \rxs\, has been linked to its glitching behaviour \citep{rea05,israel07}, but no bursts have been observed so far from this magnetar. On the other hand, \uu\ is one of the first discovered magnetars, and has been extensively monitored in the past decade. It showed a few X-ray bursts and a glitch, again with only a subtle increase in luminosity \citep{gonz10,gav11}. 

\sgra\, and \sgrb\, are the most prolific X-ray bursters among the magnetar class, and both showed a {\it giant flare}: the most energetic events ever observed from Galactic compact objects ($L\sim10^{46}$\ergs ). Despite their frequent flaring activity, only very slight flux variations have been observed from them.
In particular, \sgra\, showed an subtle increase of its burst rate and  its X-ray persistent emission during 2003 and throughout 2004, when the luminosity less than doubled with respect to the ``historical'' level \citep{mte05,woods07}. This period of intense activity culminated with a giant flare \citep{hurley05,palmer05}.
In 1998, following an intense bursting activity, a large flux increase from \sgrb\, was reported (Woods et al. 2001) using data 
from the RXTE All Sky Monitor (ASM),  but \swift\,  did not measure a large flux increase during
a similar intense bursting activity epoch in 2006  (Israel et al. 2008). The low positional and timing accuracy of the ASM, the crowded region where this source lies, and the non-detection of such large flux increase by any other 
accurate imaging instrument make us incline to wait for confirmation of such event. 
No glitch has been reported for these two sources, although the large timing noise could have hidden their glitching activity in the timing data. On the other hand, large period derivative changes have been measured, which can be the 
(non-conclusive) evidence of a missed glitches.

\subsection{Glitches and/or short bursts coincident with an outburst} 

Among the well monitored magnetar long outbursts, we have evidence of glitches and/or bursts  in several cases. \ea\, is the first long transient event discovered, and the prototype of the connection between outbursts, glitches and bursts\citep{kaspi03,woods04,zhu08}. Nevertheless, we now know that its flux variability was among the least extreme cases. \ee\, showed another episode of transient  flux increase (again not extreme though) connected with X-ray bursts and one glitch \citep{gavriil04,tmt05,tam08,dib09}. 

The most extreme transient events, corresponding with a luminosity increase of a factor of $\sim 100$ or more, have always been observed in coincidence with bursting activity, such as \xte\,  \citep{ibra04, bernardini11}, \aa\, (which showed multiple outbursts; \cite{israel10,ng10}), \wes \citep{muno07},  \sgrd\, \citep{mereghetti06,eiz08}, \sgre\, \citep{rea09}, or the newly discovered \sgrf\,  \citep{vanderhorst10,esposito10,rea10} and \sgrg\, \citep{gogus10,esposito11}. Among those extreme transients, glitches were detected only from \wes\, ($\Delta\nu / \nu > 1.5\times10^{-5}$; \cite{israel07b}; but see also \citep{woods07}).

%%%%%%%%%%% FIGURE %%%%%%%%%%%%%%%%%%%%%%%%%%%%%%%

\begin{figure*}
\includegraphics[width=8cm,height=6cm,angle=0]{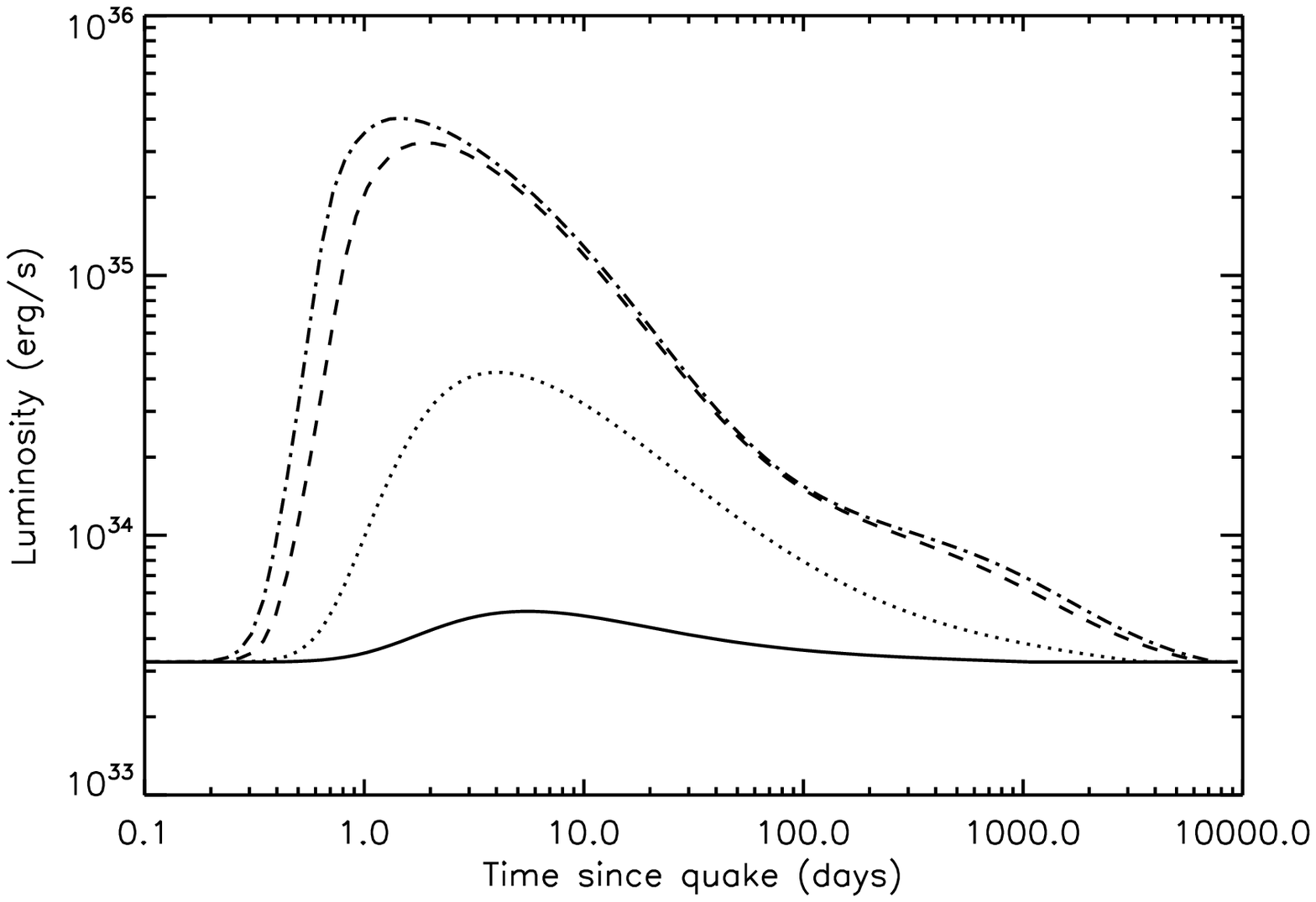}
\includegraphics[width=8cm,height=6cm,angle=0]{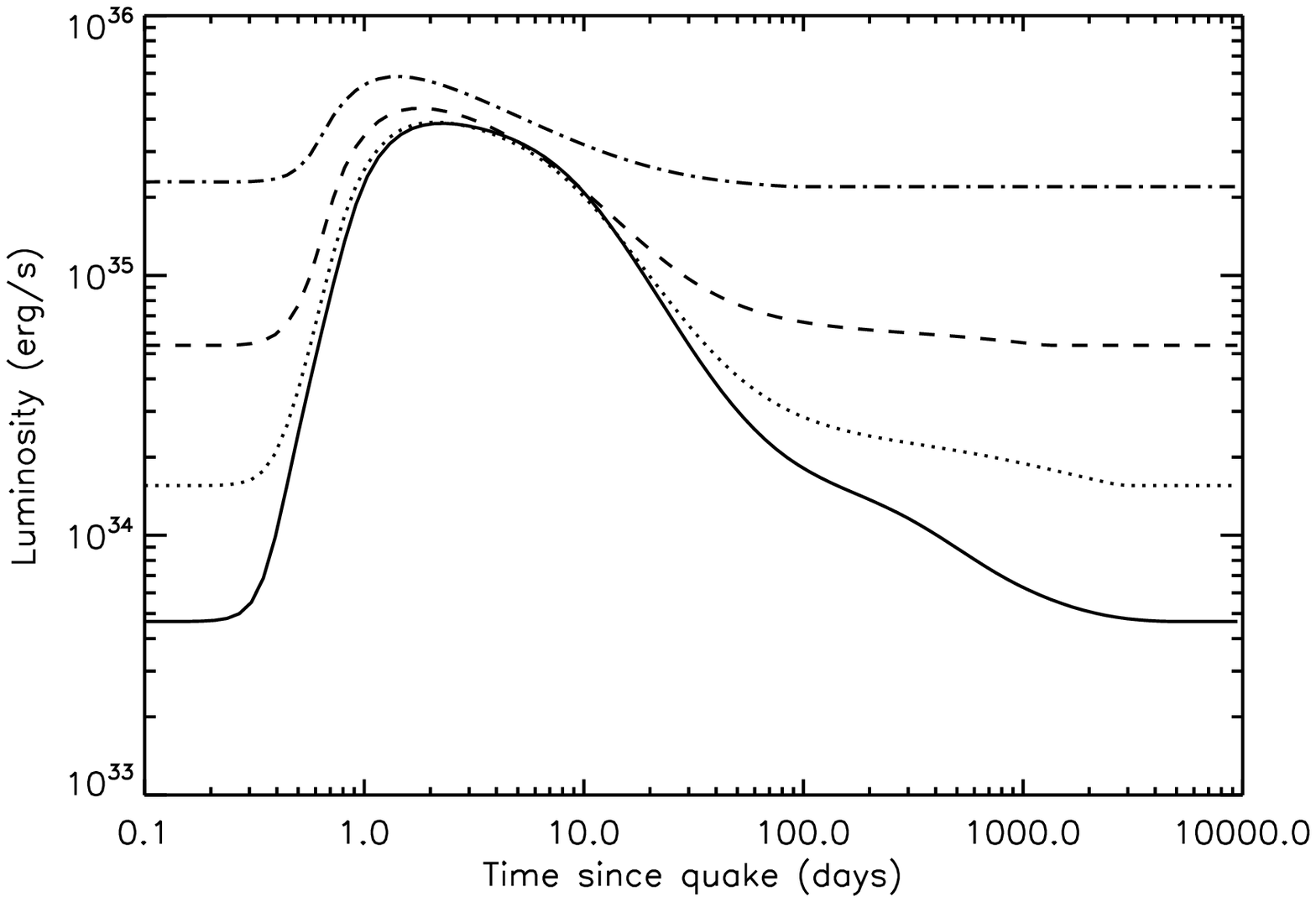}
\caption{Luminosity vs. time after energy injection. {\em Left panel}: Effect of the total energy injected. The models
correspond to $E_{\rm oc}=1.7\times10^{41}$\,erg, (solid line), $1.7\times10^{42}$\,erg (dotted line), 
 $1.7\times10^{43}$\,erg (dashed line), and $1.7\times10^{44}$\,erg (dash-dotted line). {\em Right panel}: Comparison of 
models with the same energy injection ($E_{\rm oc}=1.7\times10^{44}$ erg) 
but varying the initial state (quiescent luminosity).}
\end{figure*}

%%%%%%%%%%%%%%%%%%%%%%%%%%%%%%%%%%%%%%%%%%%%%%%%

\section{Characterizing the outburst properties and the recovery to quiescence}
\label{cooling}

Having revised the observational ground, we turn now to discuss theoretical predictions.
We use the two-dimensional  cooling code designed to study the magneto-thermal evolution of neutron stars
\citep{aguil08,pons09} to model the raise and decay of luminosity during a magnetar outburst.
We refer to these two works for technical details and the microphysics input. 
To generate an initial pre-burst model, we first evolve a standard magnetar 
($M=1.4M_\odot$,  $R =11.6$~km, $B_p=(0.5-3)\times10^{14}$ G), keeping fixed the core temperature, until a
stationary solution is obtained. By varying the core temperature and the magnetic field we control the 
surface temperature,  and therefore the quiescent luminosity of the initial model.
Once we have a starting model, we inject some fixed amount of energy in a fraction of the crust volume 
on a very short timescale (hours), and follow the evolution of the
thermal structure until it returns to the original state, typically after a few years.
On such a short timescale, the magnetic field is practically frozen.

We ran several models varying the total energy injected, the angular size, and the depth of the region 
where the energy is released. The total energy input varies in the  $10^{40}-10^{44}$\,erg range,
spanning the expected observational estimates and the theoretical predictions \citep{pern11}.
Different depths (from a thin layer to the whole crust) and angular sizes (from 0.2 radians to the entire
surface) were tried. After analyzing all the models, the most important conclusions are the following
(a longer detailed discussion will be reported in a subsequent paper of this series):

{\it 1. Dependence on the depth where the energy is injected}. We have found that nearly the totality of the energy
injected in the inner crust is efficiently radiated in the form of neutrinos, thus having no effect on the
surface temperature and the photon luminosity. This has already been noticed in previous 
one-dimensional studies \citep{Kam06}, who concluded that the heat source should be located 
at densities below the neutron drip point $\rho < 3-4 \times 10^{11}$ g cm$^{-3}$ (the outer crust), to have an impact
on the thermal luminosity. We confirm that this conclusion remains valid in 2D simulations.
Hereafter we use the energy injected in the outer crust ($E_{\rm oc}$) as the reference parameter.

{\it 2. Dependence on the angular size}. We found that angular heat transport in the outer
crust is very inefficient because, in the outer layers, the magnetic field is predominantly radial, and electron
conduction across magnetic field lines is strongly suppressed. Angular transport in the inner crust
is possible for some particular geometries but, since the energy in the inner crust is rapidly
lost by neutrino emission, this has no real effect on the photon luminosity. Hence, the size of the hot spot formed by
the energy injection remains almost constant, and only towards the end of the evolution, when
the neutron star is close to its original state, a small increase in the surface of the spot was observed.

{\it 3. Dependence on the energy injection rate.} We explored the sensitivity of our results to the variation of the 
time interval in which the energy is released (from a few minutes to one day).
The relevant parameter turned out to be again the total energy $E_{\rm oc}$, quite independently 
on the rate at which it is injected.
The injection rate affects the rise of the luminosity curve only if the energy is released very close to the surface. 
In any case, the heat wave needs some time to reach the surface, and the luminosity rise is not instantaneous
(1 hour to $1$ day), which is probably too fast to be observable.
After reaching the maximum, the cooling curve is independent on the injection rate and reflects
different physics (thermal relaxation of the crust). This happens
on a longer timescale (months to years).

{\it 4. Dependence on the total energy input.} A minimum value of $E_{\rm oc}$ is needed to 
have a visible effect. For $E_{\rm oc}<10^{40}$\,erg the event is barely observable as a slight
luminosity variation. The most relevant result is an interesting saturation effect for $E_{\rm oc}>10^{43}$\,erg. 
A larger energy release does not vary the final result. The reason for this saturation is that, 
as soon as the crust reaches $3-4 \times 10^9$\,K, neutrino processes in the outer crust are strongly 
reactivated, and the temperature 
cannot be further increased because the system self-regulates by neutrino emission. However, it should be noted that
the two most important neutrino emission processes in this regime are  plasmon and pair annihilation
(see  \cite{yak01,yp04}  for reviews on neutrino processes and neutron star cooling),  but these two processes 
in the presence of very strong magnetic fields have not been properly calculated. Further work in this line is
needed to fully understand magnetar cooling curves.

Fig. 1 (left panel), shows the temporal variation of the luminosity for four representative cases, 
varying $E_{\rm oc}$ from $10^{41}$ to $10^{44}$\,erg. In all cases heat is deposited in a region of about 200\,m depth
(between densities $\rho\sim10^8-10^{11}$\,g~cm$^{-3}$, and covering a small area of 3\% of the star surface, which corresponds to an angle of 0.5\,rad).
The delay (a few hours) between the injection of energy and the luminosity peak is caused by the time needed for 
the internal heat wave to reach the star surface.
The saturation when $E_{\rm oc}>10^{43}$\,erg is clearly visible. A larger energy release does not change the
peak luminosity, which only can be increased by enlarging the area affected.

{\it 5. Dependence on the initial state.} The other fundamental parameter to understand magnetar outbursts is the initial state. The combination of the quiescent luminosity with
the saturation effect mentioned above is crucial to understand magnetar phenomenology. 
Increasing the total energy injected does not result in higher surface temperatures, which are limited to
0.5-0.6 keV (maybe a short transient flash of a few minutes can reach slightly higher temperatures).
Therefore, the maximum thermal luminosity is also limited\footnote{Resonant comptonization in the magnetosphere can be very effective in reshaping the spectrum, but it does not
vary the total luminosity, which is fixed by the seed thermal photons from the surface. Only in the very
extreme case where most of the electrons are ultra-relativistic the luminosity can be visibly enhanced.}.
This means that, if the initial state is a very bright magnetar, the luminosity cannot be increased more than a factor of a few. On the other hand,
if the initial state consists in a dim source, we have room to increase its luminosity in 2-3 orders
of magnitude.

This is illustrated if Fig. 1 (right panel), where we compare results from different models which only differ
in the initial state (luminosity). In order to tune the luminosity of the initial state in the stationary regime,
we have varied the core temperature between $2\times10^8$ and $2\times10^9$\,K, and the
value of the poloidal field between $5\times10^{13}$ and $2.5\times10^{14}$\,G, which fixes
the heating rate by magnetic field dissipation in the crust. For simplicity, we assumed that no toroidal
field is present. The strength of the internal toroidal field is also related to the luminosity of the initial state, 
but it does not change our conclusions.  
All of the models have the same energy input: $E_{\rm oc}=1.7\times10^{44}$\,erg in the same region as before.
In the figure we can see that, for a low quiescent luminosity ($L_q = 3 \times 10^{33}$ \ergs), a starquake that
releases $\approx 10^{44}$\,erg produces an increase in the luminosity of 2 orders of magnitude in about 1 day
and its cooling curve can be followed for several years. Conversely, exactly the same type of event in a very bright
magnetar ($L_q = 3\times 10^{35}$ \ergs), is barely seen as a small variation of luminosity in a factor of 2 and lasting
only a few days.

%%%%%%%%%%% FIGURE %%%%%%%%%%%%%%%%%%%%%%%%%%%%%%%

\begin{figure}
\hspace{-1cm}
\includegraphics[width=10.4cm,height=8cm,angle=0]{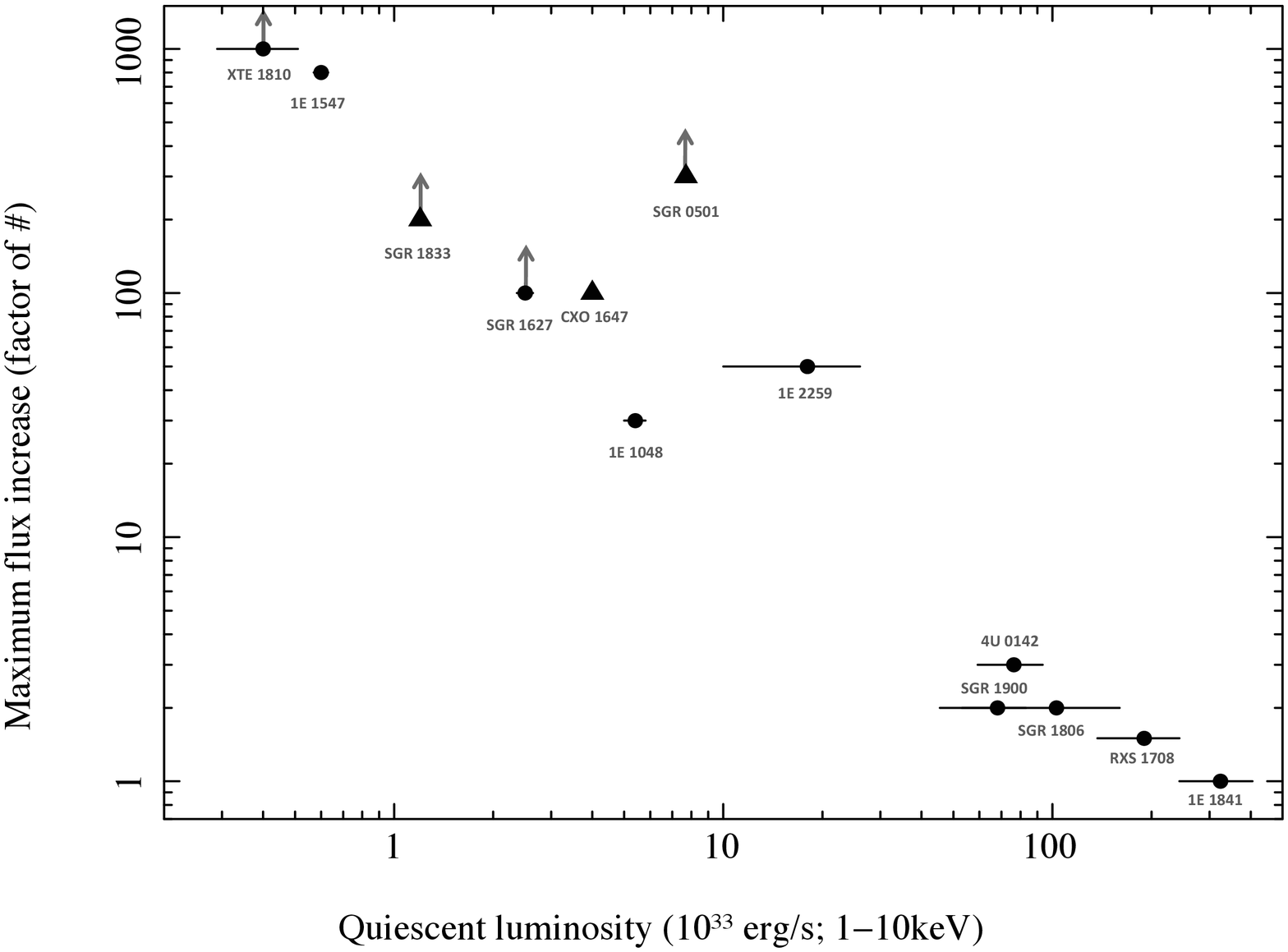}
\caption{Quiescent luminosity vs. outburst maximum flux increase (all in the 1-10 keV band), for all magnetars showing bursts, glitches or outbursts.  Errors in the measurements include the uncertainties in the flux values and in the distances.}
\end{figure}

%%%%%%%%%%%%%%%%%%%%%%%%%%%%%%%%%%%%%%%%%%%%%%%%

%%%%%%%%%%%%%%%%%%%%%%%%%%%%%
\section{Discussion}
\label{discussion}

We have discussed how the connection between outbursts, short X-ray bursts and glitches might appear rather erratic. Theoretically, glitches and short X-ray bursts are believed to be correlated to 
starquakes induced by the progressive increase of magnetic stresses in the crust. When the local conditions are such that
the system cannot stand the tension any longer, crustal fractures occur. 
They may have associated the
ejection of particles and the reorganization of the magnetosphere. At the same time or shortly after, when the heat wave 
caused by the release of energy reaches the surface, it is 
also expected the increase of the star temperature, and therefore its persistent emission. However, the lack of detection of outbursts correlated with the glitching and bursting activity of several magnetars (see \S\ref{obs}) posed several questions on the validity of this interpretation, that we now can answer.

In \S\,\ref{cooling} we studied the rise of the surface temperature of a magnetar and its subsequent cooling when
a certain amount of energy is injected in the outer crust. 
Because of the strong temperature dependence of the neutrino emission processes, 
the system can efficiently self-regulate its temperature.
The result is an upper limit to the temperature (and luminosity) at the outburst peak: even releasing
a much larger amount of energy, the luminosity will reach a maximum between $10^{35}-10^{36}$\ergs,
with the precise number depending on the area affected by the event. In other words,
any event with an energy release in the outer crust $> 10^{43}$\ergs, will show a similar maximum  outburst luminosity, regardless of their dipolar magnetic field strength, quiescent luminosity, or any other parameter involved.

In Fig.\,2 we have plotted the quiescent luminosity of all magnetars that showed glitches and/or bursts, as a function of the maximum persistent flux increase observed in each source. We only consider flux variability on timescales longer than a few days to avoid the contamination from bursts and flares. Furthermore, to select a sample as unbiased as possible, we have neglected flux variations detected with instruments with poor angular and temporal resolution which could not disentangle the contribution from single short bursts (such as RXTE-ASM, and older generation instruments). Among these, we only
consider the events for which the outburst decay was also monitored with good resolution instruments (as for \xte\, and \ea).

Although it is almost impossible to have a good quantitative estimate on how many outbursts from magnetars we might have missed in the past years, we note that since the launch of \swift\, in 2004 \citep{swift04}, we can rely on a daily coverage of the whole sky with the Burst Alert Telescope (BAT; which has a field of view of about 1/6 of the sky; 15-150\,keV), and a rapid follow-up with the \swift\, X-Ray Telescope (XRT; 0.3-10\,keV). \swift\, allowed us to collect more than a dozen of outbursts in the past 8 years, as well as to discover 5 new magnetars through their outburst activity (see Rea \& Esposito 2011 for a detailed review). This makes us relatively confident of having a good sky coverage and outbursts sample, and 
we believe that only a few events might have been missed during the \swift\, era.

Looking at Fig.\,2, a clear trend is present, with brighter objects showing less flux enhancement than dim magnetars. However, we warn that these numbers must be taken with caution due to 
1) current distance uncertainties which might well be underestimated, 2) the use of a reduced energy band of i.e. 1-10\,keV, that in combination with the spectral softening during the outburst decay can result in the underestimate of the quiescence luminosity, and 3) the uncertainty in the exact peak-flux for many of those objects.
For sources having a good pre-outburst monitoring we plot the estimate of the flux-enhancement, while we only
quote lower limits for the most uncertain cases.
In any case, all this caveats may be estimated in about factors of 2, and the correlation shown in the figure extends over
three orders of magnitude in both axis. Note also that the peak luminosity in all cases is in the expected range
of $\sim10^{35}-10^{36}$ \ergs. In particular, fitting the data in Fig.\,2 (excluding the sources for which we have only lower limits in the peak flux) we find a mean outburst peak luminosity of $\sim3.5\times10^{35}$\ergs .

The general conclusions we can extract from our results can be summarized in the following assertions:
\begin{enumerate}
\item The definition of ``transient" magnetars (AXPs or SGRs) as opposed to the so-called ``persistent magnetars'' is spurious: it only reflects their different  quiescent luminosities. 
\item Bursts and glitches are probably always accompanied by a radiative enhancement.
\item Given the same typical outburst energetics, 
large relative flux enhancements can only be observed in faint quiescent objects.
\item
Large, long flux enhancements from bright magnetars will never be observed, since their peak radiative luminosities 
cannot exceed $\sim10^{36}$\ergs, which in most cases is undetectable. At most, it may simply appear as 
subtle flux variations (as are the cases of \kes, or \rxs).
\end{enumerate}

The line dividing the historical separation between AXPs and SGRs has been erased during the last decade and now
they are thought to represent two regions of the same distribution. With the results presented here, we also show that
the same can be said for the separation between ``transient" and ``persistent" magnetars. As better data are collected
and more theoretical work is being done, the separation of magnetars in different classes according to
burst activity, timing noise, or spectral properties becomes more and more blurred.  This leads to the conclusion that
the distribution of neutron stars with relatively high magnetic fields is a continuum with no fundamental 
intrinsic separation in classes.

\acknowledgements 

This work was partly supported by Compstar, a Research Networking Programme of the European Science
Foundation and grants AYA2010-21097-C03-02, GVPROMETEO2009-13 (JAP) and
AYA2009-07391, SGR2009-811, and TW2010005 (NR).
NR is supported by a Ram\'on y Cajal fellowship through Consejo Superior de Investigaciones
Cient\'{\i}ficas (CSIC).

\end{document}